\documentclass[preprint]{elsarticle}
\usepackage[normalem]{ulem}
\usepackage{lineno,hyperref}
\modulolinenumbers[5]

\journal{Journal of \LaTeX\ Templates}

\usepackage{amsmath,amssymb}

\newcommand{\ba}{\begin{eqnarray}}
\newcommand{\ea}{\end{eqnarray}}
\newcommand{\be}{\begin{equation}}
\newcommand{\ee}{\end{equation}}

\usepackage{xcolor}

\usepackage{tikz-feynman}
\tikzfeynmanset{compat=1.1.0}
\usepackage{tikz}
\usetikzlibrary{trees}
\usetikzlibrary{decorations.pathmorphing}
\usetikzlibrary{decorations.markings}
\definecolor{MyOrange}{HTML}{FF9900}
\definecolor{MyBlue}{HTML}{3366FF}
\definecolor{MyRed}{HTML}{CC0000}

\usepackage{subcaption}

\begin{document}

\begin{frontmatter}

\title{Impact of a thermal medium on $D$ mesons and their chiral partners}

\author[1]{Gl\`oria Monta\~na}
\ead{gmontana@fqa.ub.edu}

\author[1]{\`Angels Ramos}
\ead{ramos@fqa.ub.edu}

\author[2,3,4,5]{Laura Tol\'os}
\ead{tolos@th.physik.uni-frankfurt.de}

\author[2]{Juan M. Torres-Rincon}
\ead{torres-rincon@th.physik.uni-frankfurt.de}

\address[1]{Departament de F\'isica Qu\`antica i Astrof\'isica and Institut de Ci\`encies del Cosmos (ICCUB), Facultat de F\'isica,  Universitat de Barcelona, Mart\'i i Franqu\`es 1, 08028 Barcelona, Spain}
\address[2]{Institut f\"ur Theoretische Physik, Goethe Universit\"at Frankfurt, Max von Laue Strasse 1, 60438 Frankfurt, Germany}
\address[3]{Frankfurt Institute for Advanced Studies, Ruth-Moufang-Str. 1, 60438 Frankfurt am Main, Germany}
\address[4]{Institute of Space Sciences (ICE, CSIC), Campus UAB, Carrer de Can Magrans, 08193, Barcelona, Spain}
\address[5]{Institut d'Estudis Espacials de Catalunya (IEEC), 08034 Barcelona, Spain}

\begin{abstract}
We study $D$ and $D_s$ mesons at finite temperature using an effective field theory based on chiral and heavy-quark spin-flavor symmetries within the imaginary-time formalism. Interactions with the light degrees of freedom are unitarized via a Bethe-Salpeter approach, and the $D$ and $D_s$ self-energies are calculated self-consistently. We generate dynamically the $D^*_0(2300)$ and $D_s(2317)$ states, and study their possible identification as the chiral partners of the $D$ and $D_s$ ground states, respectively. We show the evolution of their masses and decay widths as functions of temperature, and provide an analysis of the chiral-symmetry restoration in the heavy-flavor sector below the transition temperature. In particular, we analyse the very special case of the $D$-meson, for which the chiral partner is associated to the double-pole structure of the $D^*_0(2300)$.
\end{abstract}


\end{frontmatter}

% \linenumbers

\section{Introduction}

The idea that chiral partners become degenerate above the chiral restoration temperature $T_\chi$~\cite{Hatsuda:1985eb,Rapp:1999ej} has motivated a large amount of works in which low-lying hadronic states of opposite parities have been studied in a thermal medium and their masses have been seen to merge at large temperatures $T>T_\chi$.

The canonical example resides in the light-meson sector, where the pseudoscalar isotriplet ($\pi$) and the scalar isoscalar ($\sigma$ meson) acquire similar masses above $T_\chi$. This system has been studied in the linear sigma model~\cite{Bochkarev:1995gi}, the (P)NJL model~\cite{Klevansky:1992qe,Florkowski:1993br,Hansen:2006ee}, the quark-meson model~\cite{Tripolt:2013jra} and others.  On the other hand, vector and axial vector interactions, that have been studied in the (P)NJL model~\cite{Sintes:2014lka}, and gauge linear-sigma model~\cite{Pisarski:1995xu} for example, allow one to study the chiral symmetry restoration of the $\rho$ and the $a_1$ states~\cite{Rapp:1999ej}. Opposite-parity diquarks also present such degeneracy in the (P)NJL model~\cite{Torres-Rincon:2015rma}, whereas there exist also indications from lattice-QCD calculations of the chiral restoration of opposite-parity baryons~\cite{Aarts:2017rrl,Aarts:2018glk}.

In many of the theoretical models, the parity partners are fundamental degrees of freedom, e.g. $\pi$ and $\sigma$ in the linear sigma model~\cite{Bochkarev:1995gi}, and interactions in a thermal/dense medium dress them producing in-medium mass modifications. In another set of models, e.g. the NJL and PNJL model, the parity partners (either $0^+/0^-$ or $1^+/1^-$) are not part of the degrees of freedom of the Lagrangian, but are instead generated from few-body dynamics, like those implemented by the Bethe-Salpeter equation for a quark-antiquark pair. In this case, masses and decay widths seem to converge in the chirally-restored phase~\cite{Hansen:2006ee}.

All these models provide insights of the effects of chiral restoration, both below and above $T_\chi$. However one should keep in mind that---although well-motivated by the QCD symmetries and dynamics---they are not usually the correct effective field theory (EFT) of QCD. In the light-meson sector, for instance, we know that the low-energy effective theory is chiral perturbation theory (ChPT)~\cite{Gasser:1983yg}. It can lead to model-independent results, also at finite temperatures. However, this approach is valid at low energies and temperatures, always below $T_\chi$, and only timid indications of a chiral-symmetry restoration can be expected from it. 

Even if limited to $T<T_\chi$, this chiral approach is quite interesting because a combined picture of the chiral partners comes into play. The negative parity partner $\pi$ is a degree of freedom of the Lagrangian~\cite{Gasser:1983yg}, whose vacuum mass is dressed by interactions with the whole set of (pseudo-) Goldstone bosons. However, the positive parity partner ($\sigma$) is not part of the Lagrangian. In unitarized versions of ChPT~\cite{Dobado:1989qm,Dobado:1996ps} it can be associated to the $J^\pi=0^+$ resonant state, appearing in the scalar isoscalar channel of the meson-meson scattering amplitude. This state---experimentally identified with the scalar $f_0(500)$ of the Particle Data Group~\cite{Tanabashi:2018oca}---can be generated at finite temperature as well~\cite{Dobado:2002xf,Rapp:1995fv}. This scenario, where one of the chiral companions is a degree of freedom of the theory and the other a dynamically-generated state, is the one we consider in this work.

In this letter we focus on light-heavy meson systems and look for the thermal effects on the $D$ and $D_s$ mesons properties, and of their chiral partners. For this goal, we extend previous results in a more complete and consistent approach using a hadronic EFT. The spirit is similar to Ref.~\cite{Cleven:2017fun} where a chiral $SU(4)$ effective Lagrangian at leading-order (LO) was used (see also~\cite{Mishra:2003se} for a use of the same EFT). However, in the present work, we construct the interactions based on an effective Lagrangian based on $SU(3)$ chiral, and heavy-quark symmetries. This effective theory at next-to-leading-order (NLO) has been well studied in vacuum, and its low-energy parameters fixed by lattice-QCD fits \cite{Guo:2009ct,Liu:2012zya,Guo:2018tjx}. The dynamics of the light-heavy meson systems is computed at finite temperature in the framework of the imaginary-time formalism (ITF), and we use unitarity and self-consistency as our guiding principles.

An important goal of this work is to study the spectroscopy of the heavy-light sector at finite temperature. This means that we are interested in accessing not only the masses and decay widths of the $D$ and $D_s$ mesons, but also of the states which appear dynamically upon unitarization, namely the $D^*_0(2300)$ and $D_{s0}^*(2317)$ states. It happens that these scalars, positive parity states could be associated with the chiral partners of the ground states. Therefore, we can describe the temperature  dependence of their masses/widths in view of the possible restoration of chiral symmetry in heavy-light systems. Limited by low temperatures (below $T_\chi$) we simply provide qualitative indications on how these states approach the chiral transition, not being able to describe what happens above it. Moreover we  discuss a new peculiar picture of chiral companions as the $D^*_0(2300)$ is described by a double-pole structure. This is a new scenario for chiral symmetry restoration as one needs to study simultaneously the evolution with temperature of three states.

\section{Effective Lagrangian and Unitarized Interactions at $T\neq0$}

At $T < T_\chi$ and assuming no baryon density, the thermal medium is essentially composed by the lighter mesons of the pseudoscalar meson octet. Their interactions at low energies are governed by ChPT, based on chiral power counting. The heavy $J^\pi=0^-$ mesons, $D$ and $D_s$, propagate through this medium behaving as Brownian particles, suffering from collisions with any of the light mesons. The interaction of the $D$-mesons with light particles is described by an effective Lagrangian based on both chiral and heavy-quark symmetries~\cite{Kolomeitsev:2003ac,Lutz:2007sk}. We use the version at NLO in the chiral expansion, similarly as in~\cite{Guo:2009ct,Liu:2012zya,Guo:2018tjx,Geng:2010vw,Abreu2011,Tolos:2013kva,Albaladejo:2016lbb}.

The LO Lagrangian reads
\begin{align}
 \mathcal{L}_{\rm LO}&=\langle\nabla^\mu D\nabla_\mu D^\dagger\rangle-m_D^2\langle DD^\dagger\rangle-\langle\nabla^\mu D^{*\nu}\nabla_\mu D^{*\dagger}_{\nu}\rangle+m_D^2\langle D^{*\nu}D^{*\dagger}_{\nu}\rangle \nonumber \\
 & +ig\langle D^{*\mu}u_\mu D^\dagger-Du^\mu D^{*\dagger}_\mu\rangle+\frac{g}{2m_D}\langle D^*_\mu u_\alpha\nabla_\beta D^{*\dagger}_\nu-\nabla_\beta D^*_\mu u_\alpha D^{*\dagger}_\nu\rangle\epsilon^{\mu\nu\alpha\beta} \ ,
\end{align}
where $D$ denotes the antitriplet of $0^-$ $D$-mesons [$D=\begin{pmatrix} D^0 & D^+ & D^+_s \end{pmatrix}$], and similarly for the vector $1^-$ states [$D^*_\mu=\begin{pmatrix} D^{*0} & D^{*+} & D^{*+}_s \end{pmatrix}_\mu$] (not used in this work)]. The light mesons are encoded into $u_\mu=i(u^\dagger\partial_\mu u-u\partial_\mu u^\dagger)$, where $u$ is the unitary matrix of Goldstone bosons in the exponential representation. The bracket denotes the trace in flavor space and the connection of the covariant derivative $\nabla_\mu D^{(*)}=\partial_\mu D^{(*)} -D^{(*)}\Gamma^\mu$ reads $\Gamma_\mu=\frac{1}{2}(u^\dagger\partial_\mu u+u\partial_\mu u^\dagger)$.

The NLO Lagrangian is given by
\begin{align}\nonumber\label{eq:lagrangianNLO}
 \mathcal{L}_{\rm NLO}=&-h_0\langle DD^\dagger\rangle\langle\chi_+\rangle+h_1\langle D\chi_+D^\dagger\rangle+h_2\langle DD^\dagger\rangle\langle u^\mu u_\mu\rangle \\ \nonumber
 &+h_3\langle Du^\mu u_\mu D^\dagger\rangle+h_4\langle\nabla_\mu D\nabla_\nu D^\dagger\rangle\langle u^\mu u^\nu\rangle+h_5\langle\nabla_\mu D\{u^\mu,u^\nu\}\nabla_\nu D^\dagger \rangle \\ \nonumber
 &+\tilde{h}_0\langle D^{*\mu}D^{*\dagger}_\mu\rangle\langle\chi_+\rangle-\tilde{h}_1\langle D^{*\mu}\chi_+D^{*\dagger}_\mu\rangle-\tilde{h}_2\langle D^{*\mu}D^{*\dagger}_\mu\rangle\langle u^\nu u_\nu\rangle \\ 
 &-\tilde{h}_3\langle D^{*\mu}u^\nu u_\nu D^{*\dagger}_\mu\rangle-\tilde{h}_4\langle\nabla_\mu D^{*\alpha}\nabla_\nu D^{*\dagger}_\alpha\rangle\langle u^\mu u^\nu\rangle-\tilde{h}_5\langle\nabla_\mu D^{*\alpha}\{u^\mu,u^\nu\}\nabla_\nu D^{*\dagger}_\alpha\rangle,
\end{align}
where $\chi_+=u^\dagger\chi u^\dagger+u\chi u$, with the quark mass matrix $\chi={\rm diag}(m_\pi^2,m_\pi^2,2m_K^2-m_\pi^2)$. 

For more details we recommend Refs.~\cite{Geng:2010vw,Abreu2011,Liu:2012zya,Tolos:2013kva}. The low-energy constants (LECs, $h_i$ with $i=0,...,5$), have been revisited in this work in view of the recent study~\cite{Guo:2018tjx} based on lattice-QCD data.

The effective Lagrangian at LO+NLO provides the tree-level scattering amplitude for $D$ and $D_s$ mesons with light mesons,
\begin{align} \nonumber\label{eq:potential}
 V^{ij}(s,t,u)=&\frac{1}{f_\pi^2}\Big[\frac{C_{\rm LO}^{ij}}{4}(s-u)-4C_0^{ij}h_0+2C_1^{ij}h_1\\ 
 &-2C_{24}^{ij}\Big(2h_2(p_2\cdot p_4)+h_4\big((p_1\cdot p_2)(p_3\cdot p_4)+(p_1\cdot p_4)(p_2\cdot p_3)\big)\Big)\\ \nonumber
 &+2C_{35}^{ij}\Big(h_3(p_2\cdot p_4)+h_5\big((p_1\cdot p_2)(p_3\cdot p_4)+(p_1\cdot p_4)(p_2\cdot p_3)\big)\Big)
 \Big],
\end{align}
where $p_1$ and $p_2$ ($p_3$ and $p_4$) are the momenta of the incoming (outgoing) mesons and $C_{{\rm LO},0,1,24,35}$ are the isospin coefficients (see Table~II in \cite{Liu:2012zya}). The $i,j$ indices denote channels with given values of strangeness $S$ and isospin $I$.

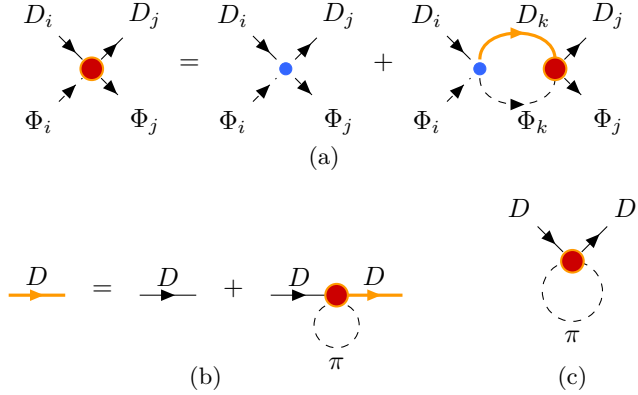
\begin{figure}[htbp!]
\centering 
\begin{subfigure}[b]{\textwidth}\centering 
\captionsetup{skip=0pt}
 \begin{tikzpicture}[baseline=(i.base)]
    \begin{feynman}[small]
      \vertex (a) {\(D_i\)};
      \vertex [below right = of a] (i) {};
      \vertex [above right = of i] (b) {\(D_j\)};
      \vertex [below right = of i] (d) {\(\Phi_j\)};
      \vertex [below left=of i] (c) {\(\Phi_i\)};
      \diagram* {
        (a) -- [fermion] (i), 
        (i) -- [fermion] (b),
        (c) -- [charged scalar] (i),
        (i) -- [charged scalar] (d),
       };
     \draw[dot,minimum size=4mm,thick,MyOrange,fill=MyRed] (i) circle(1.5mm);
    \end{feynman}
  \end{tikzpicture}
  $=$
  \begin{tikzpicture}[baseline=(i.base)]
    \begin{feynman}[small]
      \vertex (a) {\(D_i\)};
      \vertex [below right = of a] (i) {};
      \vertex [above right = of i] (b) {\(D_j\)};
      \vertex [below right = of i] (d) {\(\Phi_j\)};
      \vertex [below left=of i] (c) {\(\Phi_i\)};
      \diagram*{
        (a) -- [fermion] (i), 
        (i) -- [fermion] (b),
        (c) -- [charged scalar] (i),
        (i) -- [charged scalar] (d),
       } ;    
     \draw[dot,MyBlue,fill=MyBlue] (i) circle(.8mm);
    \end{feynman}
  \end{tikzpicture}
  $+$
  \begin{tikzpicture}[baseline=(i.base)]
    \begin{feynman}[small]
      \vertex (a) {\(D_i\)};
      \vertex [below right = of a] (i) {};
      \vertex [right = of i] (j) {};
      \vertex [above right = of j] (b) {\(D_j\)};
      \vertex [below right = of j] (d) {\(\Phi_j\)};
      \vertex [below left=of i] (c) {\(\Phi_i\)};
      \vertex [above right = of i] (b1) {\(D_k\)};
      \vertex [below right=of i] (d1) {\(\Phi_k\)};
      \diagram*{
        (a) -- [fermion] (i), 
        (j) -- [fermion] (b),
        (i) -- [MyOrange, fermion, very thick, half left, looseness=1.2] (j),
        (i) -- [charged scalar, half right, looseness=1.2] (j),
        (c) -- [charged scalar] (i),
        (j) -- [charged scalar] (d),
       } ;    
     \draw[dot,MyBlue,fill=MyBlue] (i) circle(0.8mm);
     \draw[dot,minimum size=4mm,thick,MyOrange,fill=MyRed] (j) circle(1.5mm);
    \end{feynman}
  \end{tikzpicture}
\caption{}
\label{fig:subfigA}
\end{subfigure}
\\[0.2cm]
\hspace{-1cm}\begin{subfigure}[b]{0.7\textwidth}\centering 
\captionsetup{skip=-5pt}
 \begin{tikzpicture}[baseline=(a.base)]
    \begin{feynman}[small]
      \vertex (a) {};
      \vertex [right = of a] (b) {};
      \diagram* {
        (a) -- [MyOrange, fermion, very thick,edge label=\(\textcolor{black}{D}\)] (b), 
       };
    \end{feynman}
  \end{tikzpicture}
  $=$
 \begin{tikzpicture}[baseline=(a.base)]
    \begin{feynman}[small]
      \vertex (a) {};
      \vertex [right = of a] (b) {};
      \diagram* {
        (a) -- [fermion, edge label=\(D\)] (b), 
       };
    \end{feynman}
  \end{tikzpicture}
  $+$
  \begin{tikzpicture}[baseline=(a)]
    \begin{feynman}[small, inline=(a)]
      \vertex (a) {};
      \vertex [right = of a] (i) {};
      \vertex [right = of i] (b) {};
      \vertex [below = 0.4cm of i] (d) {};
      \vertex [below = 0.9cm of i] (e) {\(\pi\)};
      \diagram* {
        (a) -- [fermion,edge label=\(D\)] (i), 
        (i) -- [MyOrange, fermion, very thick,edge label=\(\textcolor{black}{D}\)] (b),
       } ;   
     \draw[dashed] (d) circle(0.3cm);
     \draw[dot,minimum size=4mm,thick,MyOrange,fill=MyRed] (i) circle(1.5mm);
    \end{feynman}
  \end{tikzpicture}
\caption{}
\label{fig:subfigB}
\end{subfigure}\hspace{-1cm}
\begin{subfigure}[b]{0.25\textwidth}\centering
\captionsetup{skip=5pt}
  \begin{tikzpicture}
    \begin{feynman}[small]
      \vertex (a) {\(D\)};
      \vertex [below right = of a] (i) {};
      \vertex [above right = of i] (b) {\(D\)};
      \vertex [below = 0.4cm of i] (d) {};
      \vertex [below = 1cm of i] (e) {\(\pi\)};
      \diagram* {
        (a) -- [fermion] (i), 
        (i) -- [fermion] (b),
       } ;   
     \draw[dashed] (d) circle(0.4cm);
     \draw[dot,minimum size=4mm,thick,MyOrange,fill=MyRed] (i) circle(1.5mm);
    \end{feynman}
  \end{tikzpicture}
\caption{}
\label{fig:subfigC}
\end{subfigure}
\caption{(a) Bethe-Salpeter equation. The $T$-matrix is solved self-consistently with dressed internal heavy-meson propagator.  (b) Dressed heavy-meson propagator. (c) Heavy-meson self-energy. The heavy meson is dressed by the unitarized interaction with pions ($T$-matrix, red dot).}
\end{figure}

This amplitude is used as the kernel of an on-shell Bethe-Salpeter equation within a full coupled-channel basis,  $T=V+VGT$, where $T$ is the unitarized amplitude and $G$ is the light-heavy two-body propagator, which contains medium effects (see Fig.~\ref{fig:subfigA}). In IFT, after a Matsubara summation and a continuation to real energies, the loop reads
\begin{eqnarray}\label{eq:loopT}
  G_{D\Phi}(E,\vec{p};T)&=&\int\frac{d^3q}{(2\pi)^3}\int d\omega\int d\omega'\frac{S_{D}(\omega,\vec{q};T)S_{\Phi}(\omega',\vec{p}-\vec{q};T)}{E-\omega-\omega'+i\varepsilon} \nonumber \\
  && \times [1+f(\omega,T)+f(\omega',T)],
\end{eqnarray}
where $D$ denotes the heavy meson and $\Phi$ the light meson. The vacuum contribution of the above loop function needs regularization. While a dimensionally regularized loop is used in \cite{Guo:2018tjx}, with subtraction constants fitted to lattice-QCD data, we employ a cutoff scheme, a procedure which simplifies the numerical treatment at finite temperature. The value of the UV cutoff is chosen such that the loop functions in both schemes agree at threshold. To simplify further, we eliminate the channel dependence by choosing a common value of 800~MeV, which is close to the obtained values for the different channels. We checked that our results for the scattering lengths at $T=0$ are consistent with those obtained in  \cite{Guo:2018tjx}.

At $T\neq 0$ the internal meson propagators receive medium corrections due to the light meson gas. In ChPT the pion mass and decay constant do not appreciably change with temperature up to two-loops and even in unitary extensions of it~\cite{Schenk:1993ru,Toublan:1997rr}. In addition, the pion damping rate is very much suppressed at the temperatures explored in this paper, so we have decided to use the pion vacuum spectral function for all temperatures. For the $D$ meson, we consider its medium modification through a self-consistent scheme consisting in using the $T$-matrix (Fig.~\ref{fig:subfigA}) to dress the propagator (Fig.~\ref{fig:subfigB}) with the $D$-meson self-energy (Fig.~\ref{fig:subfigC}) which reads,
\begin{equation}\label{eq:selfE}
    \Pi_{D}(E,\vec{p};T)=\int\frac{d^3q}{(2\pi)^3}\int d\Omega\frac{E}{\omega_\pi}\frac{f(\Omega,T)-f(\omega_\pi,T)}{E^2-(\omega_\pi-\Omega)^2+i\varepsilon}\Bigg(-\frac{1}{\pi}\Bigg){\rm Im\,}T_{D\pi}(\Omega,\vec{p}+\vec{q};T) \ .
\end{equation}

The $D$-meson spectral function to be used in the loop function is therefore,
\begin{equation} \label{eq:specfunc}
  S_{D}(\omega,\vec{q};T)=-\frac{1}{\pi}{\rm Im\,}\mathcal{D}_{D}(\omega,\vec{q};T)=-\frac{1}{\pi}{\rm Im\,}\Bigg(\frac{1}{\omega^2-\vec{q}\,^2-m_{D}^2-\Pi_{D}(\omega,\vec{q};T)}\Bigg) \ .
\end{equation}
This set of equations is solved iteratively until self-consistency is obtained.

\section{Dynamically generated states at $T=0$}

Let us first discuss our findings at $T=0$. In order to do so, we analytically continue the energy to the complex-energy plane and look for poles in the appropriate Riemann-sheet (RS) of the $T$-matrix, to find bound, resonant and virtual states. The pole position $\sqrt{s_R}$ provides the pole mass, $M_R={\rm Re\,}\sqrt{s_R}$, and the width, $\Gamma_R/2={\rm Im\,}\sqrt{s_R}$. We also report the coupling, $|g_i|^{-2}=\partial T^{-1}_{ii} (s) /\partial s |_{s=s_R}$, of each pole to each of the channels $i$ that the pole can couple.

In this letter we focus on the sectors $(S,I)=(0,\frac12)$---with three coupled channels, viz. $D\pi(2005.3)$, $D\eta(2415.1)$ and $D_s\bar{K}(2464.0)$--- and $(S,I)=(1,0)$---with two coupled channels, $DK(2364.9)$ and $D_s\eta(2516.2)$, where the number in parentheses gives the corresponding threshold energy in MeV. We use the Fit-2B set of LECs used in \cite{Guo:2018tjx}, as it is the preferred one in that work and also the most similar to the one employed in \cite{Liu:2012zya}. Our results for the dynamically generated $0^+$ partners are summarized in Table~\ref{tab:poles}.

 \begin{table}[htbp!]
 \begin{tabular}{cccccc}
\hline
& $(S,I)$ & RS & $M_R$ & $\Gamma_R/2$ & $|g_i|$ \\ % & $X_i$ \\
& & & (MeV) & (MeV) & (GeV) \\ % &  \\
\hline
\hline
$D_0^*(2300)$ & $(0,\frac12)$ & $(-,+,+)$ & $2081.9$ & $86.0$  & $|g_{D\pi}|=8.9$ \\ %  & $X_{D\pi}=0.29-i\,0.27$ \\
 &    &           &        &       & $|g_{D\eta}|=0.4$ \\ % & $X_{D\eta}=0.00+i\,0.00$ \\
 &   &           &        &       & $|g_{D_s\bar{K}}|=5.4$ \\ % & $X_{D_s\bar{K}}=0.01+i\,0.05$ \\
 &   & $(-,-,+)$ & $2521.2$ & $121.7$ & $|g_{D\pi}|=6.4$ \\ %  & $X_{D\pi}=0.02+i\,0.09$ \\
 &   &           &        &       & $|g_{D\eta}|=8.4$ \\ % & $X_{D\eta}=0.15-i\,0.27$ \\
 &   &           &        &       & $|g_{D_s\bar{K}}|=14.0$ \\ % & $X_{D_s\bar{K}}=0.43+i\,0.49$ \\
    \hline
$D_{s0}^*(2317)$ &   $(1,0)$ & $(+,+)$ & $2252.5$ & $0.0$ & $|g_{DK}|=13.3$ \\ % & $X_{DK}=0.66+i\,0.00$ \\
 &    &           &        &       & $|g_{D_s\eta}|=9.2$ \\ % & $X_{D_s\eta}=0.17+i\,0.00$ \\
\hline
 \end{tabular}
 \centering
 \caption{Poles and the corresponding couplings to the coupled channels of the physical $D_0^*(2300)$ (first two poles) and $D_{s0}^*(2317)$ (last pole).}
 \label{tab:poles}
 \end{table}

In the sector with $(S,I)=(0,\frac12)$ we find two poles in the complex energy plane. Both correspond to the experimental $D_0^*(2300)$ state~\cite{Tanabashi:2018oca}. This double-pole structure has been previously analyzed in~\cite{Guo:2018tjx,Albaladejo:2016lbb}. The lower pole appears just above the first threshold in the $(-,+,+)$\footnote{The notation indicates the RS of the loop function for each of the coupled channels ($+$ for first and $-$ for second).} RS of the $T$-matrix around 2080 MeV. The nature of the higher pole is a bit more complicated. We find it above the $D_s\bar{K}$ threshold as a pole in the $(-,-,+)$ RS, but for some values of the parameters of the model~\cite{Guo:2018tjx,Albaladejo:2016lbb} the pole appears in the same sheet between the $D\eta$ and $D_s\bar{K}$ thresholds or below the $D\eta$ threshold\footnote{\label{footnote}We note that the $(-,-,+)$ RS is only connected to the real energy axis in the region between the $D\eta$ and the $D_s\bar{K}$ thresholds.}, strongly coupled to the $D_s\bar{K}$ channel in all cases. The dependence on the parameters is described in detail in Ref.~\cite{Guo:2018tjx}, but notice that it only brings slight modifications in the pole positions. Such an effect is not much relevant for our temperature-dependence study, as all parameters are fixed at $T=0$.

Both poles have a considerable decay width so they are not close to the real energy axis.  As we will see later, their reflection on the real axis will leave a peculiar structure, which one identifies with the experimental $D_0^*(2300)$. The lower pole couples mostly to $D\pi$, with reasonably large coupling to $D_s\bar{K}$, whereas the higher one couples to all the channels but with a larger coupling to $D_s\bar{K}$ .
 
In the $(S,I)=(1,0)$ sector the situation is somewhat clearer. We find a single pole on the real axis, which we identify to the bound state $D_{s0}^*(2317)$. It has sizable couplings to both $DK$ and $D_s \eta$, but it cannot decay to any of them as the phase space is closed at $T=0$.

\section{Spectral functions, masses, and widths at $T\neq 0$}

We now present the results of our study at finite temperature. To begin with, we dedicate a few lines commenting on the limiting temperature in our approach, apart from the evident restriction $T<T_\chi$ already mentioned. 

Light mesons composing the thermal bath are described by ChPT. According to its power counting the temperature is a soft scale contained in the expansion parameter $T/(C f_\pi)$, with $C$ a numerical factor. For example in the massless case one has $C=\sqrt{8}$~\cite{Gasser:1986vb}. In the massive case, the first works established a limiting temperature of $T\simeq 150$ MeV~\cite{Goity:1989gs}, but in practice, this value should be reduced to $T\simeq 100$ MeV, when pions start having typical energies probing the $\rho$ meson peak~\cite{Schenk:1993ru,Toublan:1997rr}. 

However the unitarized version of ChPT extends the validity of the theory to higher energies, therefore increasing the maximal temperature even above $T=150$ MeV~\cite{Dobado:2002xf,Schenk:1993ru,GomezNicola:2002tn}. In particular, the application of the same unitarization method used in this work to the light sector~\cite{Oller:1997ng} allows us to explain meson-meson scattering up to $\sqrt{s}=1.2$ GeV (see also \cite{GomezNicola:2001as} for the same conclusion using an alternative unitarization method). On the other hand, at $T=150$ MeV the mean thermal energy of a pion is $\langle E_\pi \rangle=475$ MeV, and a pair of pions colliding head-on will have a typical $\sqrt{s} \simeq 950$ MeV, which can be described by the unitarized theory. For a $D-\pi$ head-on collision one has a typical $\sqrt{s} \simeq 2.6$ GeV, which is near the higher pole of the $D^*_0(2300)$ resonance, and therefore within the validity of our theory. While for kaons the typical thermal energies are higher and can exceed $\sqrt{s}=1.2$ GeV, a generic collision will not occur head-on and the typical $\sqrt{s}$ decreases below the limit. In conclusion, we determine that $T=150$ MeV should be a reasonable limiting temperature for our theory. Of course, one should take the results around this value with caution, because the fastest mesons might lie outside the validity of the theory and, in addition, the deconfined phase will start playing a role in the system.

\begin{figure}[htbp!]
\centering
\includegraphics[scale=0.5]{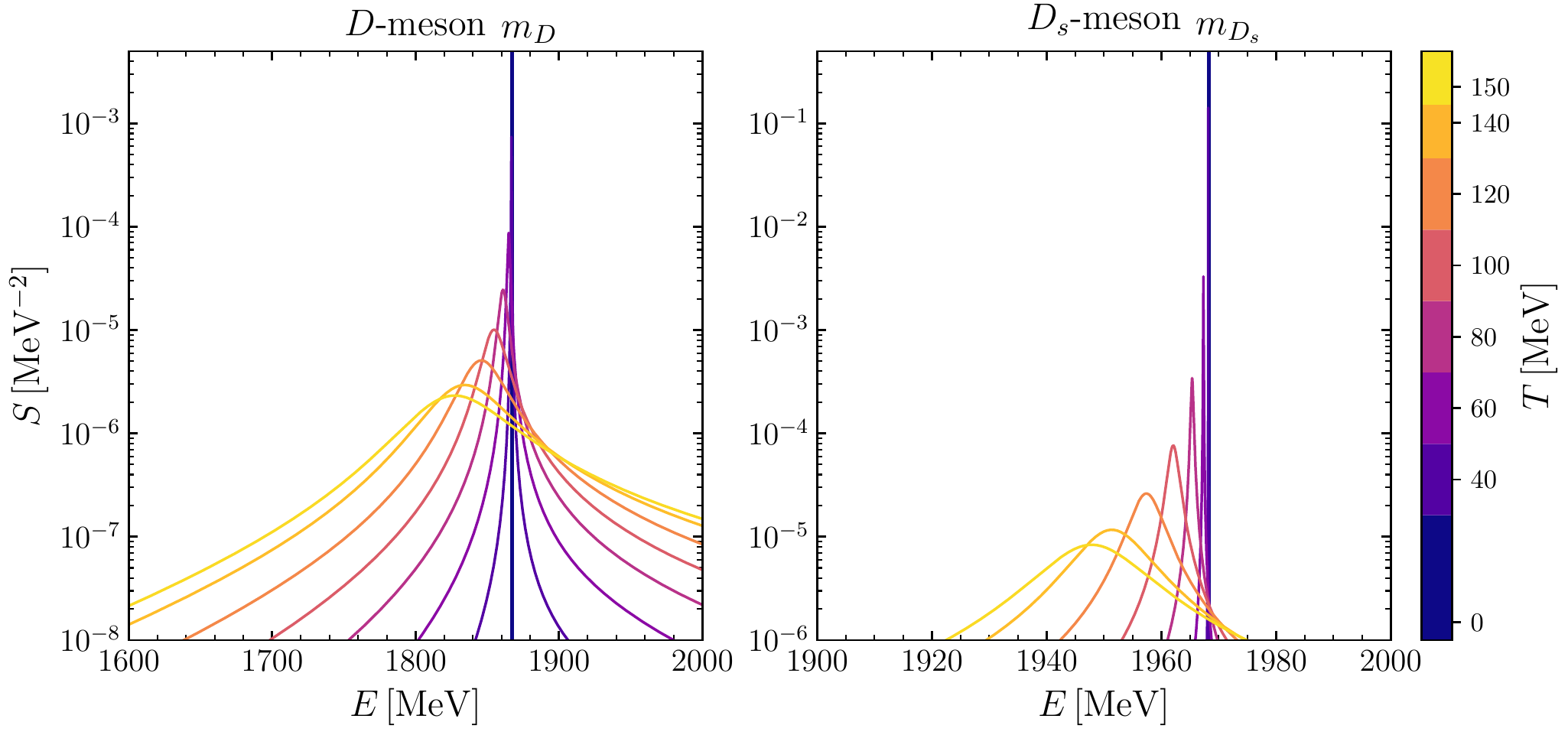}
\includegraphics[scale=0.5]{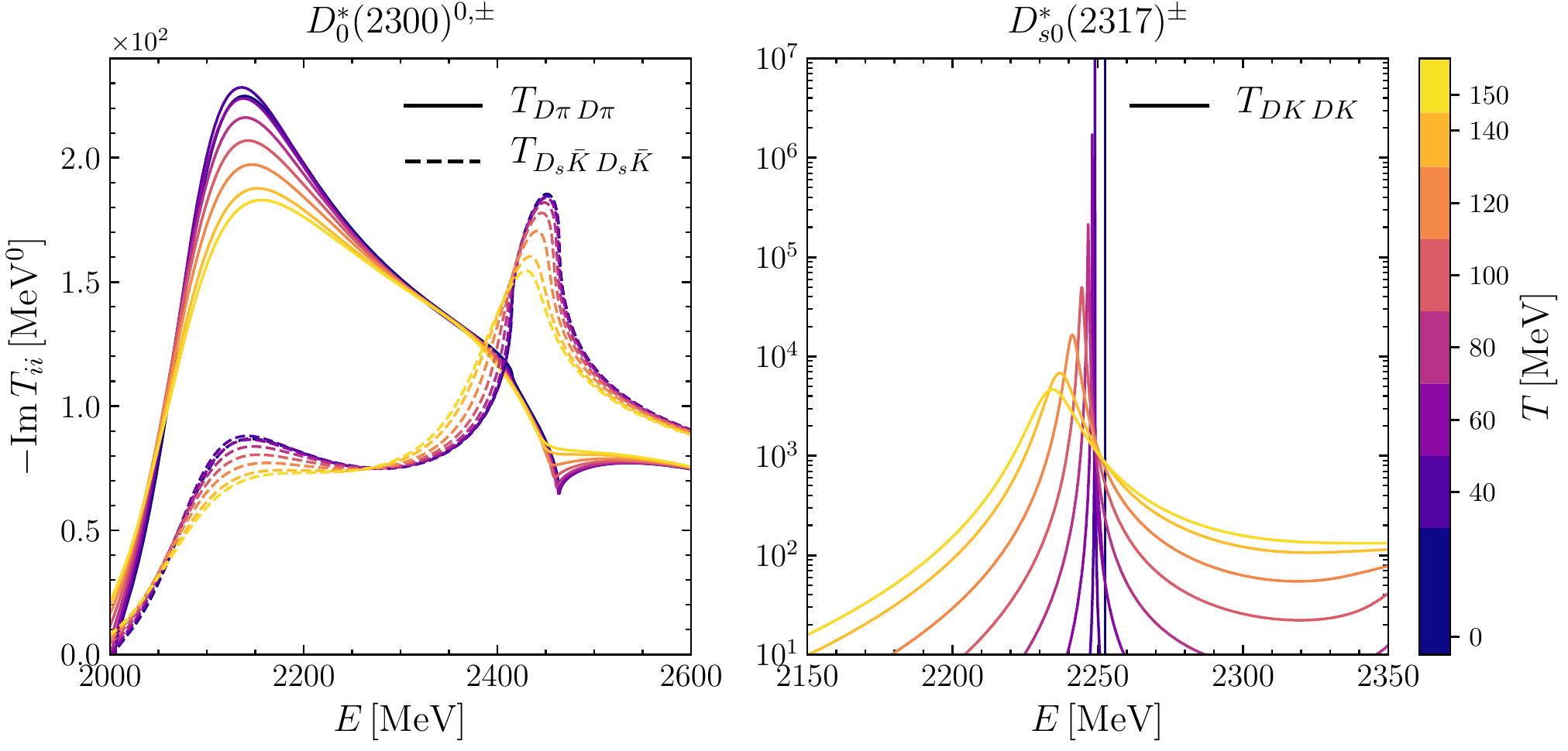}
\caption{Upper plots: spectral functions of the $D$- (left) and $D_s$-mesons (right) at different temperatures from 0 to 150 MeV. Lower plots: imaginary part of the $D\pi\rightarrow D\pi$ and $D_s\bar{K}\rightarrow D_s\bar{K}$ scattering amplitudes in the $(S,I)=(0,\frac12)$ sector (left)  and the $DK\rightarrow DK$ amplitude for $(S,I)=(1,0)$ (right) at the same values of the temperature.}
\label{fig:spectral}
\end{figure}

The spectral functions of the $D$ and $D_s$ mesons follow the standard definition in terms of the retarded propagator, see Eq.~(\ref{eq:specfunc}). They are shown on the top panels of Fig.~\ref{fig:spectral} at zero trimomentum, as functions of the energy and for different temperatures (coloured lines). The mass shift and widening of both states with temperature is evident, being these effects stronger for the $D$-meson, whose mass decreases considerably with $T$. The properties of the dynamically generated states are directly obtained from the imaginary part of the amplitudes $T_{ii}$ as a proxy for their spectral shape. It is presented in the bottom panels of Fig.~\ref{fig:spectral}, with $i$ denoting the channel to which the state couples most, i.e. $D\pi$ ($D_s \bar{K}$) for the lower (higher) pole of the $D_0^*(2300)$ in the $(S,I)=(0,\frac12)$ sector, and $DK$ for the pole of the $D_{s0}^*(2317)$ in the $(S,I)=(1,0)$ sector. In the $S=0$ case peculiar structures appear, which are produced by  the interplay of the position of the resonance to some nearby channel thresholds. Still the evolution of the peak and width of the amplitudes with $T$ is evident. For the $S=1$ sector the situation is clearer, but one can observe that, in addition to the typical thermal widening, more strength is visible on the right-hand side tail producing a totally asymmetric distribution. The reason lies in the fact that the unitary $DK$ threshold is lowered due to the decrease of the $D$ mass and its widening with temperature, hence opening the phase space for decay into this channel at smaller energies.

\begin{figure}[htbp!]
\centering
\includegraphics[scale=0.6]{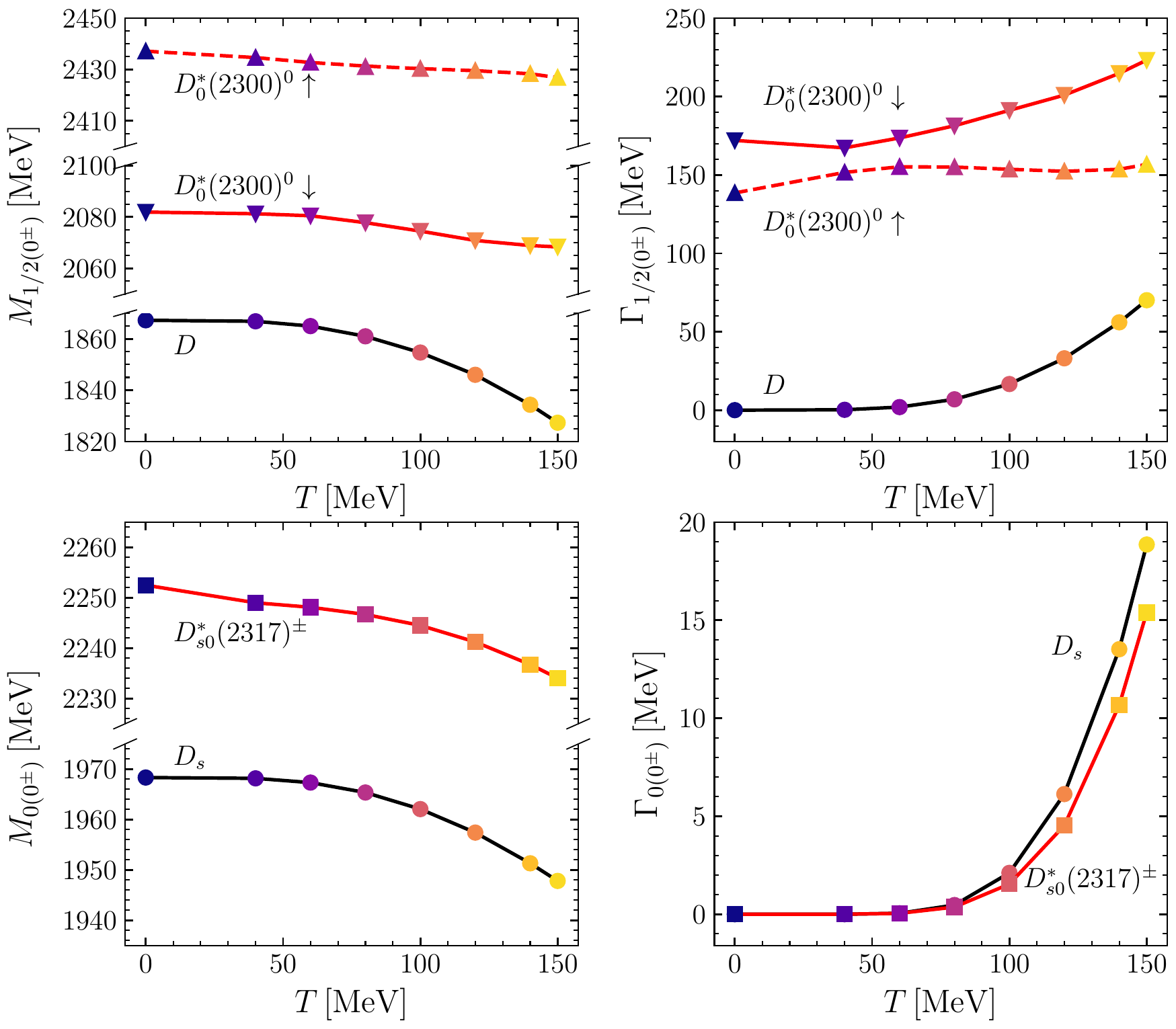}
  \caption{Temperature evolution of the mass (left panels) and width (right panels) of the chiral partners in the $(S,I)=(0,\frac12)$ sector (upper panels) and in the $(S,I)=(1,0)$ sector (lower panels). The ground-state $0^-$ partners are represented with circles and the dynamically generated $0^+$ partners, the two poles of the $D_0^*(2300)$ and the $D_{s0}^*(2317)$ pole, with upward/downward triangles and squares, respectively.}
  \label{fig:masseswidths}
\end{figure}

Finally in Fig.~\ref{fig:masseswidths} we represent the evolution of the masses and decay widths with temperature. Differently from the $T=0$ case, the results of which we presented in Table~\ref{tab:poles}, we find the determination of the poles in the complex energy plane unfeasible. Apart from complications tied to the analytic continuation of imaginary frequencies to the different RSs, a numerical search on the complex plane within self-consistency is computationally challenging.

Therefore, the mass and width will be obtained from the position and the half-width at half-maximum of the peak of the spectral functions in the real-energy axis. For the ground states, $D$ and $D_s$, this method is totally acceptable as the quasi-particle approximation is entirely justified. However, for the dynamically generated states---at least in the $S=0$ channel---this entails more problems because their poles are located far from the real axis and the width is not a well-defined concept. In view of these problems, we establish the following strategy, the details of which will be given in a subsequent publication,

\begin{itemize}
 \item For the lower resonance in the $(S,I)=(0,\frac12)$ sector we assume a Breit-Wigner-Fano shape~\cite{Fano:1961zz}, which takes into account the interaction between the resonance and the background corresponding to the higher resonance. The mass and width of the fit at $T=0$ are in very good agreement with the values of the pole mass and the width in Table~\ref{tab:poles}.
 \item For the higher resonance in the $(S,I)=(0,\frac12)$ sector we subtract the background contribution of the lower resonance and then fit a Flatt\'e-type distribution that describes the shape of resonances in the proximity of a threshold~\cite{Flatte:1976xu}, extended here to the three coupled-channel case. We note that this fitting procedure is very sensitive to fitting details. We present here the results for the masses and widths with fitting parameters constrained by the behaviour of the T-matrix around that resonance as seen in the lower left panel of Fig.~\ref{fig:spectral}. We defer a thorough study of the uncertainties tied to the fitting procedure to a subsequent publication. 
 \item For the resonance in the $(S,I)=(1,0)$ sector we again fit a Breit-Wigner-Fano distribution, although a simple fit with a Breit-Wigner gives the same results for $T<120$ MeV.
\end{itemize}

From the results in Fig.~\ref{fig:masseswidths} and in comparison with previous works, we list the following observations:

\begin{enumerate}
 \item The ground state $D$ mass has a sizable decrease of $\Delta m_D \sim 40$ MeV at the highest temperature $T=150$ MeV. This reduction is consistent, albeit twice larger, with that observed in~\cite{Fuchs2006}, where a more phenomenological approach is used to compute the $D$-meson propagator. Our reduction, on the other hand, is smaller than the one reported in Ref.~\cite{Sasaki2014}, that uses non-unitarized ChPT. However, in the $SU(4)$ effective approach of \cite{Cleven:2017fun} no significant modification is reported. In our present work the two poles of the $D^*_0(2300)$ have a more stable trend. They slightly move downwards, moderately distancing from each other. Therefore, in this sector we cannot conclude that masses of opposite parity states become degenerate close to $T_\chi$, although the temperatures studied might be still low for the chiral symmetry restoration.  In~\cite{Buchheim2018}  a large reduction in the mass of the positive-parity $D$ meson partner, of around 150~MeV, is found at $T=150$ MeV, but using a constant $D$ mass as an input of the sum-rule analysis. An even larger reduction of close to 200 MeV is seen in the results of~\cite{Sasaki2014}.
 
 \item The width of all states in the non-strange sector increases with temperature. The ground state shows a width of around $\sim70$ MeV at $T =150$ MeV, consistent with~\cite{Cleven:2017fun} and the estimates of Refs.~\cite{Fuchs2006,He:2011yi}. The widths of the two poles of the $D_0^*(2300)$ obtained from the fits increase moderately with temperature with respect to their vacuum value.

 \item In the strangeness sector we observe a clearer picture. The parity partners seem to decrease their mass with temperature, in a similar amount for both states, reaching a reduction of $\sim 20$ MeV for the $0^-$ state and $\sim 19$ MeV for the $0^+$ state at $T=150$ MeV. Consequently, they are still far from chiral degeneracy. These behaviours seem to be compatible with the low temperature trends seen in the linear-sigma model calculation of~\cite{Sasaki2014}.

 \item The decay widths of both strange partners increase from zero at similar rates. The width of the $D_{s0}^* (2317)$  is comparable to that acquired by the $D_s$ ground state at $T=150$ MeV. We note that, whereas the width of the latter is only due to medium effects, the $D_{s0}^* (2317)$ also contains the additional contribution of the decay into $DK$ states due to the reduction of the mass and the widening of the $D$-meson. We are not aware of any previous result to compare to in this sector.

\end{enumerate}

Apart from the above comparisons with previous models, unfortunately there is no solid data from first principles to compare to. However, in spite of the limitations in obtaining reliable information from finite temperature lattice QCD simulations tied to the difficulties in extracting the spectral function from the lattice correlators, we can still aim at a qualitative comparison. We note that a recent lattice-QCD calculation~\cite{Kelly:2018hsi} presents the spectral functions of $D$ and $D_s$ channels at different temperatures. The analysis in that paper concludes that no medium modification with respect to the $D$ and $D_s$ ground states is seen up to $T_{\chi} $, where  $T_ {\chi} \simeq 185$ MeV in that work. Given the precision of the lattice-QCD data this might be well in agreement with our findings here, as our $D$ ($D_s$) mass shift is only $2 \%$  ($1\%$) of the mass itself. As a pion mass of $m_\pi \sim 380$ MeV is used in~\cite{Kelly:2018hsi},  it would be interesting to re-address our calculation with a heavier pion mass and analyze the effects on the charm meson properties for temperatures $T<T_{\chi} $.

\section{Conclusion}

In this letter we report our findings on the properties of heavy-light mesons at finite temperatures. Using a thermal effective field theory based on chiral and heavy-quark symmetries at NLO, and on the basis of unitarized scattering amplitudes and self-consistency, we have obtained the temperature dependence of the spectral functions of the chiral partners, $D$ and $D_0^*(2300)$, as well as those of the $D_s$ and $D_s^*(2317)$ mesons.

From these spectral functions, we have extracted the dependence of the masses and widths of the mesons with temperature. In the $(S,I)=(0,\frac12)$ and $(S,I)=(1,0)$ sectors we do not observe a clear tendency to chiral degeneracy. However, we are limited by the low-temperature application of the hadron effective theory and, from the results of effective models in the light sector~\cite{Florkowski:1993br,Hatsuda:1994pi}, such degeneracy might occur at higher temperatures, $T>T_\chi$.

One of our main results is that the chiral partner of the $D$ meson, the $D_0^*(2300)$, has a double-pole structure in the complex-energy plane, and it is unclear at this point how the chiral symmetry restoration should be realized. Will both poles merge into a single one before becoming degenerate with the ground state? Or will only one pole survive and become degenerate with the ground state at $T>T_\chi$, while the other follows a different path?

Finally, we should mention that these results are important for a realistic analysis of heavy-ion collisions using appropriately medium-modified properties and/or heavy-flavor transport coefficients~\cite{Tolos:2013kva,Ozvenchuk:2014rpa,Song:2015sfa,Tolos:2016slr,Das:2016llg}.  This is mandatory to understand the mechanisms of charm production and properly characterise the deconfined and hadronic phases. We plan to address studies in that direction in the future.

\section{Acknowledgements}

J.M.T.-R. acknowledges the hospitality of the Institut de Ci\`encies de l'Espai (CSIC) and the Universitat de Barcelona, where part of this work was carried out. He thanks discussion with \'A. G\'omez-Nicola and J.A. Oller on the subject.

G.M. and A.R.  acknowledge  support from the Spanish Ministerio de Econom\'ia y Competitividad (MINECO) under the project MDM-2014-0369 of ICCUB (Unidad de Excelencia ``Mar\'ia de Maeztu''), and, with additional European FEDER funds, under the contract FIS2017-87534-P. G.M. also acknowledges support from the FPU17/04910 Doctoral Grant from MINECO. L.T. acknowledges support from the FPA2016-81114-P Grant from Ministerio de Ciencia, Innovaci\'on  y  Universidades,  Heisenberg  Programme  of the Deutsche Forschungsgemeinschaft (DFG, German research Foundation) under the Project Nr. 383452331 and THOR COST Action CA15213. L.T. and J.M.T.-R. acknowledge support from the DFG through projects no. 411563442 (Hot Heavy Mesons) and no. 315477589 - TRR 211 (Strong-interaction matter under extreme conditions).

%\section*{References}

\bibliography{D-mesonChiralLetter}

\end{document}